\documentclass[aps,prb,twocolumn,amsmath,amssymb,superscriptaddress,citeautoscript,longbibliography,floatfix]{revtex4-1}

\usepackage{graphicx}
\usepackage{dcolumn}
\usepackage{bm}
\usepackage{color}
\usepackage{mathptmx}
\usepackage{mathrsfs}
\usepackage{multirow}
\usepackage{natbib}

\begin{document}

\title{Pressure-tuning of the electrical-transport properties in the Weyl semimetal TaP }

\author{M.\ Besser}
\affiliation{Max Planck Institute for Chemical Physics of Solids, N\"{o}thnitzer Str.\ 40, 01187 Dresden, Germany}

\author{R.\ D.\ dos Reis}
\affiliation{Max Planck Institute for Chemical Physics of Solids,  N\"{o}thnitzer Str.\ 40, 01187 Dresden, Germany}
\affiliation{Brazilian Synchrotron Light Laboratory (LNLS), Brazilian Center for Research in Energy and Materials (CNPEM),Campinas, Sao Paulo, Brazil}

\author{F.-R.\ Fan}
\affiliation{Max Planck Institute for Chemical Physics of Solids,  N\"{o}thnitzer Str.\ 40, 01187 Dresden, Germany}

\author{M.\ O.\ Ajeesh}
\affiliation{Max Planck Institute for Chemical Physics of Solids,  N\"{o}thnitzer Str.\ 40, 01187 Dresden, Germany}

\author{Y.\ Sun}
\affiliation{Max Planck Institute for Chemical Physics of Solids,  N\"{o}thnitzer Str.\ 40, 01187 Dresden, Germany}

\author{M.\ Schmidt}
\affiliation{Max Planck Institute for Chemical Physics of Solids,  N\"{o}thnitzer Str.\ 40, 01187 Dresden, Germany}

\author{C.\ Felser}
\affiliation{Max Planck Institute for Chemical Physics of Solids,  N\"{o}thnitzer Str.\ 40, 01187 Dresden, Germany}

\author{M.\ Nicklas}\email{Michael.Nicklas@cpfs.mpg.de}
\affiliation{Max Planck Institute for Chemical Physics of Solids,  N\"{o}thnitzer Str.\ 40, 01187 Dresden, Germany}

\date{\today}

\begin{abstract}

We investigated the pressure evolution of the electrical transport in the almost compensated Weyl semimetal TaP. In addition, we obtained information on the modifications of the Fermi-surface topology with pressure from the analysis of pronounced Shubnikov-de Haas (SdH) quantum oscillations present in the Hall-effect and magnetoresistance data. The simultaneous analysis of the Hall and longitudinal conductivity data in a two-band model revealed an only weak decrease in the electron- and hole charge-carrier densities up to 1.2~GPa, while the mobilities are essentially pressure independent along the $a$-direction of the tetragonal crystal structure. Only weak changes in the SdH frequencies for $B\parallel a$ and $B\parallel c$ point at a robust Fermi-surface topology. In contrast to the stability of the Fermi-surface topology and of the density of charge carriers, our results evidence a strong pressure variation of the magnitude of transverse magnetoresistance for $B\parallel a$ contrary to the results for $B\parallel c$. We can relate the former to an increase in the charge-carrier mobilities along the crystallographic $c$-direction.

\end{abstract}

\maketitle

\section{INTRODUCTION}

A Weyl semimetal (WSM) can be considered a three dimensional analog of graphene in terms of its electronic dispersion. The Weyl fermions realize several exotic properties in materials, such as the presence of open Fermi surfaces and extraordinary transport properties, which makes them interesting for both basic science and technological applications \cite{Hosur:2013eb,Potter:2014cc,binghai_ARCM_2017,Huang2015,Binghai-review:2012,Baum2015,Qi2015MoTe2,Sun2015MoTe2,Soluyanov:2015WSM2,Zhang2015quantum,Du:2015TaP,Sun2015arc,Shekhar2015}.  {\color{black}One of the most striking characteristics of the WSM is the chiral or Adler-Bell-Jackiw anomaly, a chirality imbalance in the presence of parallel magnetic and electric fields, which  is expected to induce a negative longitudinal magnetoresistance (MR) \cite{Arnold,Reis_NJP,Parameswaran2014,Nielsen1983,Hirchberger_2016,Kim2013,Zhang2015ABJ}. Although a negative longitudinal MR  was detected in several putative WSMs and interpreted as a manifestation of the chiral anomaly, it is well known that for the realization of the chiral anomaly the chirality needs to be well-defined, \textit{i.e.}, the Fermi energy has to stay close enough to the Weyl nodes\cite{Nielsen1983}. This is only the case for few of the materials. Therefore, it is of fundamental interest to find ways of tuning materials to well defined chirality to realize the chiral anomaly effect.}

TaP belongs to the TaAs-family of compounds, with the other three members TaAs, NbP, and NbAs, which represents the first realization of WSMs \cite{Liu2015NbPTaP,Arnold,Shekhar2015,Yang2015TaAs,Xu2015TaAs,Lv2015TaAsbulk,Sun2015arc,Moll2015}. The interest in this family was especially triggered by the discovery of a huge nonsaturated MR and an ultrahigh carrier mobility in NbP \cite{Shekhar2015}. TaP crystallizes in a non-centrosymmetric tetragonal lattice structure with mirror symmetry in the $(100)$ and $(010)$ directions. The broken inversion symmetry of the crystal structure gives rise to 12 Weyl-point pairs, 8 pairs called W2 sitting at the top and bottom of the Brillouin zone and 4 pairs called W1 at half height at the zone boundaries. Each pair is separated by one of the mirror planes. Arnold \textit{et al.}\ used quantum oscillation data in combination with ab-inito band structure calculations to establish a picture of the Fermi surface of TaP \cite{Arnold}. The experiments revealed that the Weyl points sit energetically slightly below and above the Fermi energy $E_F$, with the W1 points off by $-41$~meV and the W2 points being closer to $E_F$ at just 13~meV above \cite{Arnold}. The W2 points are separated by a 16~meV barrier along the connecting line of the pair. In particular, they do not sit in independent Fermi-surface pockets, which makes the chirality not well-defined in TaP \cite{Arnold}.

Application of pressure is a powerful tool to tune the electronic properties of materials without introducing additional disorder. It promises, in principle, a way to vary the energies of the Weyl points and/or to modify the Fermi-surface topology. Therefore, it might be possible to change the Fermi-surface topology in such a way that the W2 points sit in independent Fermi-surface pockets implying a well defined chiral anomaly. Since surface-sensitive probes such as angle-resolved photoemission studies cannot be used for detecting the topological states under pressure, quantum oscillation investigations become the only tool to determine the effects of pressure on the Fermi-surface topology here. Electrical transport data provide, in addition to the information on the Fermi-surface extracted from Shubnikov-de Haas (SdH) oscillations, access on the density and mobility of the charge carriers. So far, there have been only a few studies on the effect of pressure on the electrical-transport properties of the TaAs-family, NbAs \cite{Luo2016NbAs,Zhang2015NbAs}, NbP \cite{Reis_PRB,Einaga17,Gupta18}, and TaAs \cite{Zhou2015TaAs}. In all these studies the crystalline and electronic structure are shown to be very stable.

\section{METHODS}

High-quality single crystals of TaP were grown via a chemical vapor transport reaction. More details on the sample preparation and characterization can be found in Ref.\ \onlinecite{Arnold}. The electrical-transport experiments were performed on high quality TaP single crystals in magnetic fields up to $B=9$~T and temperatures down to $T=1.4$~K in a $^{4}$He cryostat (JANIS) equipped with a superconducting magnet. The electrical contacts to the sample were prepared with $25~\mu{\rm m}$ platinum wires which were spot welded to the samples. Hydrostatic pressure was generated using a clamp-type pressure cell utilizing silicon oil as pressure transmitting medium. The pressure inside the cell was determined by measuring the shift of the superconducting transition temperature of a piece of Pb. Two pressure cells were set up: in the first one the magnetic field was applied parallel to the crystallographic $c$-axis and in the second one parallel to the $a$-axis. The electrical current was always flowing along one of the $a$-axes of the tetragonal crystal structure and perpendicular to the direction of the magnetic field.

The density-functional calculations were performed by using the Vienna Ab-initio Simulation Package \cite{kresse1996}. The core electrons were represented by the projector-augmented-wave potential \cite{Bloechl94,Kresse1999}, and the Perdew-Burke-Ernzerhof generalized-gradient approximation was employed for the exchange functional \cite{Perdew96}. The plane-wave energy cutoff is 350~eV. An $11\times11\times11$ $k$-point grid was used for total-energy integral in the whole Brillouin zone. After fully optimized, the force of all atoms is smaller than 0.01 eV/${\rm \AA}$ for different volumes. The corresponding pressures were derived by fitting with the Murnaghan's equation of state \cite{Fu1983}.

\section{RESULTS}

\begin{figure}[tb!]
\begin{center}
 \includegraphics[width=0.95\linewidth]{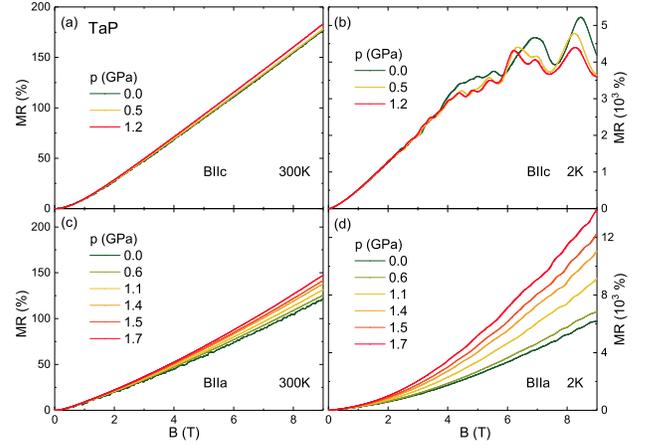}
  \caption{Transverse magnetoresistance of TaP for the two measurement configurations (a) and (b) $B\parallel c$ and (c) and (d) $B\parallel a$ at room temperature and 2K, respectively. In all experiments the electric current was applied perpendicular to the magnetic field and in parallel to the crystallographic $a$-axis.}\label{TaP_MR}
\end{center}
\end{figure}

The field dependence of the transverse MR, defined as ${\rm MR}(B)=[\rho_{xx}(B)-\rho_{xx}(0)]/{\rho_{xx}(0)}$, at 300~K and 2~K is depicted in Fig.~\ref{TaP_MR}. Here, $\rho_{xx}$ is the resistivity recorded with current and magnetic field perpendicular to each other. We performed MR measurements under magnetic fields up to 9~T and to pressure up to 1.2 and 1.7~GPa for $B\parallel c$ and $B\parallel a$, respectively. At ambient pressure we find a good agreement with the previously reported data \cite{Arnold}. Owing to the high charge-carrier mobility in TaP \cite{Arnold}, we observe a large and unsaturated MR for both field orientations at all temperatures between 2 and 300~K. However, there are differences in the behavior of the MR under pressure between the two orientations. For $B\parallel c$, the magnitude of the MR remains almost unaltered under pressure in the complete temperature range. Only the oscillatory part displays characteristic changes which we will discuss below. The robustness of the amplitude of MR$_{B\parallel c}$ in TaP is similar to previous observations for other monopinictides Weyl compounds \cite{Reis_PRB,Luo2016NbAs,Zhang2015NbAs}. On the other hand, for $B\parallel a$ only small oscillations are visible in large magnetic fields, but the application of pressure induces a continuous increase of the amplitude of the MR in the entire temperature range.  MR$_{B\parallel a}({\rm9~T})$ increases by about 21\% at 300~K and more than doubles its value at 2~K between ambient pressure and 1.7~GPa. We can argue that the large MR observed in TaP along the crystallographic $a$-direction for both field directions is directly related to the almost perfect balance between electron- and hole-like charge carriers \cite{Arnold}. The large pressure-induced increase in MR$_{B\parallel a}$ could be caused by a modification in the mobility of the electron- and/or hole-charge carriers which may originate from the anisotropic Fermi surface of TaP {\color{black} consisting of only two different types of banana shaped electron and hole pockets} \cite{Arnold} or may be related to a shift of the W2-Weyl points toward the Fermi energy.

First we will present a detailed discussion of the Hall- and magnetoconductivities, which give access to the density and mobility of the charge carriers and their correlation with the observed changes in the MR. For a sample with tetragonal crystal symmetry with magnetic field applied along the crystallographic $c$-axis, the longitudinal and transverse (Hall) components of the conductivity tensor $\sigma_{xx}$ and $\sigma_{xy}$ can be calculated from the longitudinal $\rho_{xx}$ and transverse $\rho_{xy}$ resistivities measured along the crystallographic $c$- and $a$-axes, respectively, as
\begin{equation}
\sigma_{xx} = \frac{\rho_{xx}}{\rho_{xx}^2 + \rho_{xy}^2}
\end{equation}
and
\begin{equation}
\sigma_{xy} = \frac{-\rho_{xy}}{\rho_{xx}^2 + \rho_{xy}^2}{\rm .}
\end{equation}

\begin{figure}[tb!]
	\begin{center}
		\includegraphics[width=0.95\linewidth]{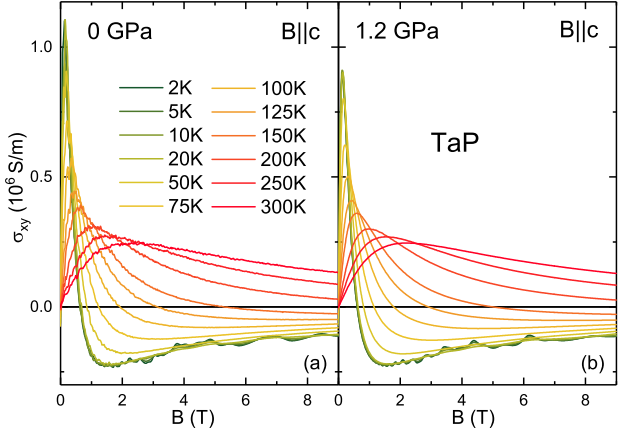}
		\caption{Magnetic-field dependence of the Hall conductivity $\sigma_{xy}$ in the $B\parallel c$ configuration for TaP for different  temperatures at (a) ambient pressure and (b) at 1.2~GPa. }\label{TaP_BIIc_sigma_xy_vs_B}
	\end{center}
\end{figure}

The non-trivial shape of the Hall conductivity as function of the magnetic field obtained for TaP, as displayed in Fig.\ \ref{TaP_BIIc_sigma_xy_vs_B}, is typical for a material with high mobilities and a nearly perfect compensation of electron- and hole-like charge carriers. The Hall conductivities for 0 and 1.2 GPa at different temperatures are displayed in Figs.\ \ref{TaP_BIIc_sigma_xy_vs_B}a and \ref{TaP_BIIc_sigma_xy_vs_B}b, respectively. The characteristic changes of $\sigma_{xy}(B)$ with temperature are almost the same for both pressures. The observed behavior is typical of a change from dominant electron-like charge transport at low temperatures to hole-like charge transport at high temperatures \cite{Shekhar2015,Arnold}.

\begin{figure}[tb!]
	\begin{center}
		\includegraphics[width=0.95\linewidth]{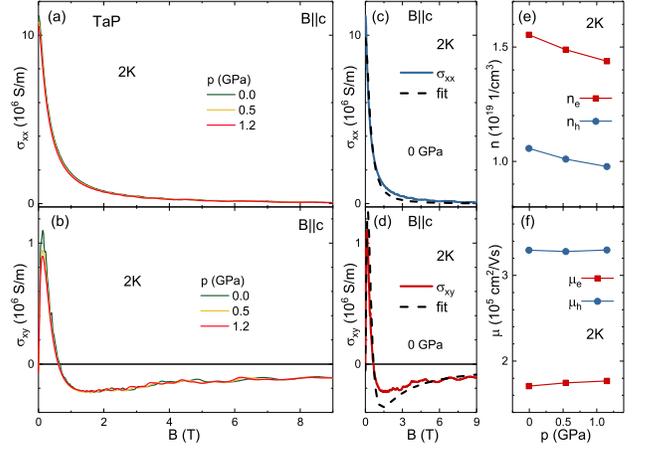}
		\caption{(a) Longitudinal electrical conductivity $\sigma_{xx}$ and (b) Hall conductivity $\sigma_{xy}$ of TaP for $B\parallel c$ at 0, 0.5, and 1.2~GPa at 2~K. (c) $\sigma_{xx}$ and (d) $\sigma_{xy}$ and the corresponding fitting curves following a two-band model at 0~GPa and 2~K. See text for details. Pressure dependencies of the (e) carrier density of electrons $n_e$ and holes $n_h$ and (f) the mobility of electrons $\mu_e$ and holes $\mu_h$ along the crystallographic $a$-direction obtained from the fits. }\label{TaP_Bc_sigma_m_n}
	\end{center}
\end{figure}

In the following, we will focus on the pressure effect on the Hall conductivity. The pressure evolution of $\sigma_{xx}(B)$ and $\sigma_{xy}(B)$ at 2~K can be seen in Figs.\ \ref{TaP_Bc_sigma_m_n}a and \ref{TaP_Bc_sigma_m_n}b, respectively. The only major change upon increasing pressure, apart from the modifications in the oscillating parts of the conductivities at magnetic fields above $1.5$~T, is in the reduction of size of the peak at small magnetic fields, in both $\sigma_{xx}(B)$ and $\sigma_{xy}(B)$.

Such a non-trivial shape of $\sigma_{xx}(B)$ and $\sigma_{xy}(B)$ can be described in a two-band Drude model including electrons and holes:
\begin{equation}
\sigma_{xy} = eB\Bigg(\frac{n_h \mu_h^2}{1+(\mu_hB)^2} - \frac{n_e \mu_e^2}{1+(\mu_eB)^2}\Bigg) {\rm ,}
\label{eqn_sigma_drude}
\end{equation}
with $n_e$ ($n_h$) and $\mu_e$ ($\mu_h$) being the carrier density and mobility for the electrons (holes). A fit of Eq.\ \ref{eqn_sigma_drude} to the Hall-conductivity data does not converge reliably and depends strongly on the initial parameters. Therefore, we decided to fit the $\sigma_{xy}(B)$ and the $\sigma_{xx}(B)$ data simultaneously using Eq.\ \ref{eqn_sigma_drude} and
\begin{equation}
\sigma_{xx} = e\Bigg(\frac{n_h \mu_h^2}{1+(\mu_hB)^2} + \frac{n_e \mu_e^2}{1+(\mu_eB)^2}\Bigg)
\label{eqn_sigma_xx_drude}
\end{equation}
for the longitudinal conductivity $\sigma_{xx}$.  The simultaneous fits converge fast independent of the initial parameters.
Representative results for $\sigma_{xx}(B)$ and $\sigma_{xy}(B)$ at 0~GPa and 2~K are displayed in Figs.\ \ref{TaP_Bc_sigma_m_n}c and \ref{TaP_Bc_sigma_m_n}d. The fitting curves describe the data at all pressures quite well.

The pressure dependencies of the carrier density and of the mobility along the crystallographic $a$-direction for both, electron- and hole-like charge carriers, are displayed in Figs.\ \ref{TaP_Bc_sigma_m_n}e and \ref{TaP_Bc_sigma_m_n}f. At ambient pressure, we find a reasonably good agreement of our carrier density and mobility data with literature \cite{Arnold}. $n_e$ and $n_h$ obtained from our data differ by about 50\% which is slightly larger than previously reported at ambient pressure \cite{Arnold}. We note that Arnold \textit{et al.} \cite{Arnold} used a different fitting procedure and did not fit $\sigma_{xx}(B)$ and $\sigma_{xy}(B)$ simultaneously. This might explain the observed small differences. Upon application of pressure, $n_e(p)$ and $n_h(p)$ start to decrease; $n_e(p)$ at a slightly larger rate than $n_h(p)$. The mobilities are almost pressure independent. Only $\mu_e(B)$ shows a slight trend to increase upon increasing pressure.

\begin{figure}[tb!]
\begin{center}
  \includegraphics[width=0.95\linewidth]{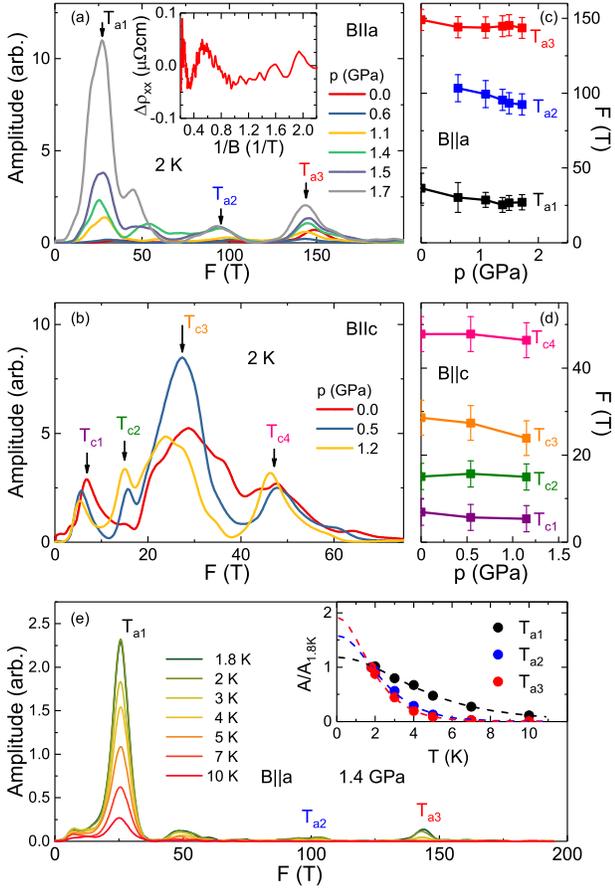}
  \caption{Fast Fourier transform of the Shubnikov-de Haas oscillations as a function of the inverse magnetic field taken at different pressures for (a) $B\parallel a$ and (b) $B\parallel c$, respectively. The oscillatory part, inset in (a), was extracted by a subtraction of a third-order polynomial background. Pressure dependencies of the SdH oscillation frequencies for (c) $B\parallel a$ and (d) $B\parallel c$, respectively. {\color{black}The error bars have been estimated taking into account the accuracy in determining the position of the maximum and the width of the peaks. (e) Fast Fourier transforms for $B\parallel a$ at 1.4 GPa for different temperatures. Inset: temperature dependence of the amplitudes. The dashed lines are fits of the Lifshitz-Kosevich formula to the data.} }\label{TaP_SdH_A-F}
\end{center}
\end{figure}

To pinpoint the pressure evolution of the Fermi-surface topology we analyze the pronounced SdH oscillations in the MR. For both magnetic field orientations, $B\parallel a$ and $B\parallel c$, SdH oscillation are clearly visible starting from about 1~T indicating very small effective masses, resulting in high mobilities (see Figs.~\ref{TaP_MR}b and \ref{TaP_MR}d). The oscillatory part of the signal $\Delta\rho_{xx}$ was obtained by subtracting a third-order polynomial background from the MR data. The inset of Fig.\ \ref{TaP_SdH_A-F}a presents $\Delta\rho_{xx}(1/B)$ at 2~K for 0 GPa as an example.  As expected, we find $\Delta\rho_{xx}$ periodic in $1/B$. The SdH frequencies were then determined by a fast Fourier transformation on the oscillatory part of the signal (see Figs.~\ref{TaP_SdH_A-F}a and \ref{TaP_SdH_A-F}b).

In the $B\parallel a$ configuration three frequencies can be extracted at ambient pressure: $T_{a1}=(36\pm10)$~T, $T_{a2}=(103\pm9)$~T and $T_{a3}=(149\pm7)$~T. These frequencies match well with the hole-pocket frequencies $F_\gamma=34$~T and $F_{\gamma '}=105$~T and the electron-pocket frequency $F_{\delta}=147$~T found in a previous ambient pressure study on TaP \cite{Arnold}. For the frequencies obtained for $B||c$, $T_{c2}=(15\pm3)$~T, $T_{c3}=(28\pm3)$~T, and $T_{c4}=(48\pm4)$T, we also find a quite reasonable correspondence to the hole pockets $F_\beta=18$~T and $F_\gamma=25$~T, and $F_\delta=45$~T originating from the electron pockets \cite{Arnold}.
In general, the SdH frequencies change little with increasing pressure. Except $T_{c2}$, which is rather stable, all frequencies decrease slightly upon application of pressure. 

We can exclude a frequency shift due to a slight tipping of the sample inside the pressure cell upon changing pressure, which would alter the angle between the crystal and the magnetic field axes and, thereby, change the extremal orbit probed by the oscillations. The angular dependence of the SdH frequencies at ambient pressure indicate that a tipping away from the $B\parallel a$ orientation would lead to a decrease of $T_{a1}$ but a simultaneous increase of $T_{a2}$ \cite{Arnold}. Similar opposite shifts should be expected for frequencies with $B\parallel c$ \cite{Arnold}. Both, we did not observe in our data. This makes a movement of the sample in the pressure cell very unlikely. We point out that beside the fundamental frequencies, a series of their higher harmonics is also present, even more pronounced in the spectra at higher pressures, manifesting the high sample quality.

While the frequencies are almost not affected, application of external pressure leads to pronounced changes in the amplitudes of the quantum oscillations. The amplitudes of the oscillations are directly related to the curvature $|\partial^{2}A/\partial k_{\parallel}^{2}|$ of the corresponding Fermi-surface cross sections. This implies that the change of the amplitudes of the SdH oscillations is caused by  modifications in the curvature close to an extremal cross section of the Fermi surface. Thus, our results point at {\color{black}tiny} modifications of the Fermi-surface {\color{black}geometry} while the overall shape is robust against the application of pressure.

{\color{black} The analysis of the temperature dependence of the SdH amplitudes provides information on the effective masses. Figure~\ref{TaP_SdH_A-F}e displays the fast Fourier transforms for different temperatures at 1.4~GPa for $B\parallel a$. By fitting the Lifshitz-Kosevich formula to the data we obtain the effective masses (see inset of Fig.~\ref{TaP_SdH_A-F}e) \cite{Lishitz56}. For the main frequencies at 1.4~GPa and $B\parallel a$, $m^*_{T_{a1}}=(0.19\pm0.01)m_0$, $m^*_{T_{a2}}=(0.39\pm0.03)m_0$, and $m^*_{T_{a3}}=(0.58\pm0.02)m_0$. Here $m_0$ is the mass of the free electron. We recognize that the effective masses are significantly enhanced upon application of pressure compared with the values at ambient pressure, $m^*_{T_{a1}}=0.021m_0$, $m^*_{T_{a2}}=0.35m_0$, and $m^*_{T_{a3}}=0.4m_0$ \cite{Arnold}.
}
\section{DISCUSSIONS}

The analysis of the SdH oscillations proves the robustness of the Fermi-surface topology of TaP for pressures up to 1.7~GPa, with the frequencies being nearly unaffected by pressure. Similar results have been also found for NbP \cite{Reis_PRB}. However, the amplitudes of some frequencies for $B\parallel a$ are strongly reduced upon increasing pressure, while the changes for the other frequencies for both $B\parallel a$ and $B\parallel c$ are rather small. On the other hand, upon application of pressure the transverse MR for $B\parallel a$ increases strongly, by 100\% at 9~T and a pressure of 1.7~GPa, while it is nearly unchanged in magnitude for $B\parallel c$.

The strong pressure-induced changes in the MR$_{B\parallel a}$ point at pronounced changes in the carrier mobilities along the crystallographic $c$-direction. In general, in high purity semimetals the balance between electron- and hole-type charge carrier densities and a high charge-carrier mobility may lead to a highly enhanced MR. From our analysis of the Hall and longitudinal conductivities ($B\parallel c$) we know that the application of pressure basically leaves the balance between electron and hole densities unchanged. Furthermore, the electron and hole mobilities along the crystallographic $a$-direction are pressure independent. Therefore, we may conclude that the strong enhancement of the transverse MR$_{B\parallel a}$ is caused by an increase of the mobilities along the $c$-axis related to {\color{black}tiny} modifications of the Fermi-surface {\color{black}geometry} as we will discuss below.

\begin{figure}[tb!]
\begin{center}
   \includegraphics[width=0.95\linewidth]{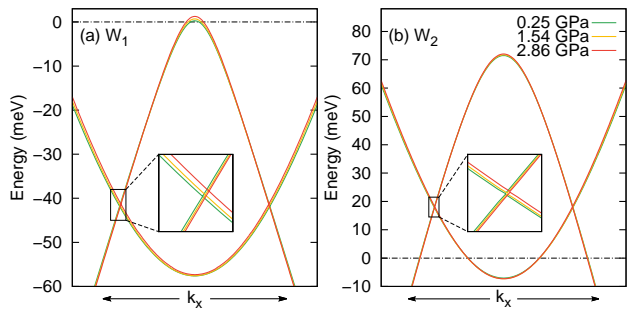}
   \caption{Band structure along the connecting line between a pair of Weyl points for (a) W1 and (b) W2. The Fermi energy is set to zero and marked by a dashed line.}\label{TaP_DOS_Weyl}
\end{center}
\end{figure}

We can obtain additional information on the energy difference between the Weyl points and the Fermi energy from our density-functional calculations. We select three different pressures in the investigated pressure range and shifted the Fermi energy of the band structures around the W1- and W2-type Weyl points to zero. The position of both types of Weyl points remains almost unchanged with pressure as shown in Fig.~\ref{TaP_DOS_Weyl}. This is in contrast to the results on NbP where the W2-type Weyl points move considerably toward the Fermi energy with increasing pressure \cite{Reis_PRB}.

It is, though, not likely that the observed enhancement of the MR$_{B\parallel a}$ is related to Weyl physics, since our band-structure calculations do only show minute modifications of the position of the Weyl points with pressure. On the other hand, the systematic drop of the amplitudes of mainly the $T_{a1}$ and, to a lesser extent, of the $T_{a3}$ frequencies can be attributed to subtle modifications in the shape of the Fermi-surface in the vicinity of the extremal orbits. Note, $T_{a1}$ and $T_{a3}$ correspond to electron and hole orbits, respectively \cite{Arnold}. {\color{black} As the analysis of the temperature dependence of the SdH amplitudes shows,} the changes of the Fermi-surface topography cause {\color{black} an increase} in the effective mass of the respective charge carriers and, therefore, in their mobility which in consequence could raise the MR for ${B\parallel a}$.

\section{CONCLUSIONS}

To summarize, in the noncentrosymmetric Weyl semimetal TaP the transverse magnetoresistance along the crystallographic $a$-axis for $B\parallel a$ is strongly enhanced by application of pressure, while the magnitude of MR$_{B\parallel c}$ is almost unaffected. Furthermore, our study shows the stability of the electronic structure in TaP which is evidenced by the pressure insensitivity of the SdH frequencies. While the overall topology of the Fermi-surface is stable against pressure small modifications of the topography in the vicinity of some of the extremal orbits is suggested by strong pressure dependencies of the magnitudes in the fast Fourier transform spectra. The analysis of the Hall- and magnetoconductivities yields almost unchanged charge-carrier mobilities along the crystallographic $a$-direction and pressure-independent densities of the electron and hole charge carriers up to 1.2~GPa. Along the $c$-direction we infer pronounced rise in charge-carrier mobilities, which can explain the strong enhancement of the transverse MR$_{B\parallel c}$ along the crystallographic $a$-axis.

\begin{acknowledgments}
We thank F.\ Arnold, E.\ Hassinger, and M.\ Naumann for stimulating discussions. R.\ D.\ dos Reis acknowledges financial support from the S\~{a}o Paulo Research Foundation (FAPESP) by the Grant No.\ 2018/00823.
\end{acknowledgments}

\bibliography{TaP-references}

\end{document}